\documentclass[final]{IEEEtran}

\usepackage[encapsulated]{CJK}
\usepackage{ucs}
\usepackage[utf8x]{inputenc}
\usepackage[cmex10]{amsmath}
\usepackage{amsmath,amssymb,amscd,bbm,amsthm,mathrsfs,dsfont}
\usepackage{algorithmic,algorithm}
\usepackage{mdwmath}
\usepackage{mdwtab}
\usepackage{bm,upgreek}
\usepackage{cite}
\usepackage{rotating,graphics,psfrag,epsfig}
\usepackage{array}
\usepackage{booktabs}
\usepackage{indentfirst}
\usepackage{subfigure}
\usepackage{lipsum,fancyhdr,lastpage,refcount}
\usepackage{url}
\usepackage{blindtext}
\usepackage[T1]{fontenc}
\usepackage{color}
\usepackage{stfloats}

\IEEEoverridecommandlockouts

\let\oldnl\nl
\newcommand{\nonl}{\renewcommand{\nl}{\let\nl\oldnl}}

\begin{document}

\title{Decentralized Deep Reinforcement Learning for Delay-Power Tradeoff in Vehicular Communications}

\author{\IEEEauthorblockN{Xianfu Chen, Celimuge Wu, Honggang Zhang, Yan Zhang, Mehdi Bennis, and Heli Vuojala}

\thanks{X. Chen and H. Vuojala are with the VTT Technical Research Centre of Finland, Finland (email: \{xianfu.chen, heli.vuojala\}@vtt.fi). C. Wu is with the Graduate School of Informatics and Engineering, University of Electro-Communications, Japan (email: clmg@is.uec.ac.jp). H. Zhang is with the College of Information Science and Electronic Engineering, Zhejiang University, China (e-mail: honggangzhang@zju.edu.cn). Y. Zhang is with the Department of Informatics, University of Oslo, Norway (e-mail: yanzhang@ieee.org). M. Bennis is with the Centre for Wireless Communications, University of Oulu, Finland (email: mehdi.bennis@oulu.fi).}

%

\vspace{-0.5cm}
}

\maketitle

\begin{abstract}

This paper targets at the problem of radio resource management for expected long-term delay-power tradeoff in vehicular communications.
At each decision epoch, the road side unit observes the global network state, allocates channels and schedules data packets for all vehicle user equipment-pairs (VUE-pairs).
The decision-making procedure is modelled as a discrete-time Markov decision process (MDP).
The technical challenges in solving an optimal control policy originate from highly spatial mobility of vehicles and temporal variations in data traffic.
To simplify the decision-making process, we first decompose the MDP into a series of per-VUE-pair MDPs.
We then propose an online long short-term memory based deep reinforcement learning algorithm to break the curse of high dimensionality in state space faced by each per-VUE-pair MDP.
With the proposed algorithm, the optimal channel allocation and packet scheduling decision at each epoch can be made in a decentralized way in accordance with the partial observations of the global network state at the VUE-pairs.
Numerical simulations validate the theoretical analysis and show the effectiveness of the proposed online learning algorithm.

\end{abstract}

\section{Introduction}
\label{intr}

The vehicle-to-vehicle (V2V) communication technologies have been gaining increasing popularity for the feasibility of enabling emerging vehicle-related services \cite{Kuut18, Dai1900, Zhang18}.
However, this ad hoc type of vehicular communications requires intense coordinations among the vehicles in close proximity \cite{Amad16}.
Without the support of an infrastructure, the high vehicle mobility makes the design of efficient radio resource management (RRM) techniques extremely challenging \cite{Zhen15}.
There are a large body of literatures on RRM in V2V communications.
In \cite{Sun16}, Sun et al. proposed a separate resource block and power allocation algorithm for the RRM in device-to-device based V2V communications.
In \cite{Yao17}, Yao et al. derived a loss differentiation rate adaptation scheme to meet the stringent delay and reliability requirements for V2V safety communications.
In \cite{Egea16}, Egea-Lopez et al. designed a fair adaptive beaconing rate algorithm for the problem of beaconing rate control in inter-vehicular communications.
Most of these efforts have not taken into account the network dynamics, such as the temporal and spatial variations in transmission quality as well as data traffic, and hence fail to optimize the expected long-term RRM performance.

A Markov decision process (MDP) has been successfully applied to model RRM in vehicular communications with time-varying nature.
In \cite{Liu18}, Liu and Bennis formulated a latency and reliability \cite{Meh18} constrained transmit power minimization problem, for which the Lyapunov stochastic optimization was leveraged to handle the network dynamics.
The problem with the Lyapunov stochastic optimization is that only an approximately optimal solution can be constructed.
In \cite{Chen18S}, Chen et al. studied the non-cooperative RRM in vehicular communications from an oblivious game-theoretic perspective and put forward an online algorithm based on reinforcement learning to approach the optimal solution.
Consider a more practical scenario, where the channel qualities are affected by the vehicle mobility, the explosion in the state space makes the technique developed in our priori work \cite{Chen18S} infeasible.

\begin{figure}[t]
  \centering
  \includegraphics[width=18pc]{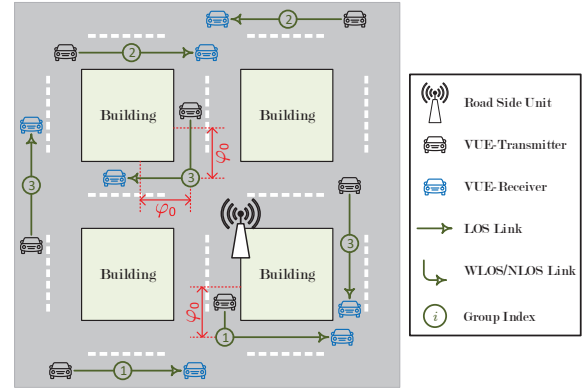}
  \caption{An illustrative Manhattan grid vehicle-to-vehicle communication network (VUE: vehicle user equipment; LOS: line-of-sight; WLOS: weak-line-of-sight; NLOS: non-line-of-sight.).}
  \label{systModeFigu}
\vspace{-0.6cm}
\end{figure}

In this paper, we investigate a Manhattan grid V2V network, where the data traffic changes across the time horizon and the channel quality state depends on the locations of vehicle user equipment (VUE)-transmitter (vTx) and VUE-receiver (vRx) of a VUE-pair.
The primary goal of this paper is to design an optimal RRM algorithm for each VUE-pair to strike a tradeoff between the queuing delay and the transmit power consumption over the long run.
We formulate the RRM problem as a MDP and resort to a deep neural network based function approximator to deal with the curse of state space explosion \cite{Mnih15}.
In \cite{Ye18M}, Ye and Li devised a decentralized RRM mechanism based on deep reinforcement learning (DRL) for V2V communication systems.
However, the mechanism does not account for the vehicle mobility, which helps facilitate frequency resource sharing among different groups of VUE-pairs.
As the major contribution from this paper, we propose an online decentralized learning algorithm by exploring the recent advances in both long short-term memory (LSTM) \cite{Hoch97} and DRL \cite{Hass16}, with which each VUE-pair with partially local network state observations is hence able to realize a significant performance improvement.

\section{System Model}
\label{systMode}

As in Fig. \ref{systModeFigu}, we consider a Manhattan grid V2V communication scenario.
A set $\mathcal{K} = \{1, \cdots, K\}$\footnote{For a well defined road segment, the VUE density tends to be steady \cite{Zhua12}.} of VUE-pairs share a set $\mathcal{J} = \{1, \cdots, J\}$ of orthogonal channels within the coverage $\mathcal{C}$ of a road side unit (RSU), where $\mathcal{C}$ represents a two-dimensional Euclidean space.
The time horizon is discretized into decision epochs, each of which is of duration $\delta$ and is indexed by an integer $t \in \mathds{N}_+$.
Each vTx always follows the corresponding vRx with a fixed distance of $\varphi$ and the vRx moves in $\mathcal{C}$ according to a Manhattan mobility model \cite{Chen18S}.
Denote by $\mathbf{x}_k^t = (x_{k}^{(1), t}, x_k^{(2), t})$ and $\mathbf{y}_k^t = (y_k^{(1), t}, y_k^{(2), t})$, respectively, the Euclidean coordinates of the vTx and the vRx of a VUE-pair $k \in \mathcal{K}$ during each epoch $t$.
Depending on whether the vTx and the vRx are in the same lane or in perpendicular lanes, the channel model during each decision epoch belongs to: 1) line-of-sight (LOS) -- both the vTx and the vRx are in the same lane; 2) weak-line-of-sight (WLOS) -- the vTx and the vRx are in perpendicular lanes and at least one of them is near the intersection within a distance of $\varphi_0$; and otherwise, 3) none-line-of-sight (NLOS).
More specifically, the channel quality state $g_{k, j}^t = \nu_{k, j}^t \cdot H_k^t \in \mathcal{G}$ experienced by VUE-pair $k$ over channel $j \in \mathcal{J}$ during epoch $t$ includes a fast fading component $\nu_{k, j}^t$ of a Rayleigh distribution with a unit scale parameter and a path loss $H_k^t$ that applies the model in (\ref{pathLossMode}) for urban areas using 5.9 GHz carrier frequency \cite{Liu18},
\begin{figure*}[!t]
\vspace{-0.2cm}
\begin{align}\label{pathLossMode}
 H_k^t =
 \left\{\!\!
 \begin{array}{l@{~}l}
   \rho \cdot \left(\sqrt{\left|x_k^{(1), t} - x_k^{(2), t}\right|^2 + \left|y_k^{(1), t} - y_k^{(2), t}\right|^2}\right)^{-e},
                                                                                                        &\mbox{if VUE-pair $k$ is in LOS}        \\
   \rho \cdot \left(\left|x_k^{(1), t} - x_k^{(2), t}\right| + \left|y_k^{(1), t} - y_k^{(2), t}\right|\right)^{-e},
                                                                                                        & \mbox{if VUE-pair $k$ is in WLOS}      \\
   \xi  \cdot \left(\left|x_k^{(1), t} - x_k^{(2), t}\right| \cdot \left|y_k^{(1), t} - y_k^{(2), t}\right|\right)^{-e},
                                                                                                        & \mbox{if VUE-pair $k$ is in NLOS}
 \end{array}
 \right.
\end{align}
\vspace{-0.2cm}
\hrule
\vspace{-0.5cm}
\end{figure*}
where $e$ is the path loss coefficient while $\rho$ and $\xi$ are the path loss exponents with $\xi < \rho \cdot (\varphi_0 / 2)^e$.

In order to mitigate the interference during wireless transmissions and maximize the channel utilization, the RSU clusters\footnote{Considering the vehicle mobility, clustering is done every $T$ epochs \cite{Liu18}.} the VUE-pairs into a set $\mathcal{I} = \{1, \cdots, I\}$ of disjoint groups based on their geographical locations, where $I > 1$.
The RSU allocates $J$ channels to the $I$ groups, while in each group, we assume that a VUE-pair can be assigned at most one channel and a channel can be assigned to at most one VUE-pair.
Let $\mathbf{u}_k^t = (u_{k, j}^t: j \in \mathcal{J})$ denote the channel allocation for a VUE-pair $k \in \mathcal{K}_i$ during decision epoch $t$, where $\mathcal{K}_i$ is the set of VUE-pairs in a group $i \in \mathcal{I}$ and
\begin{align}
   u_{k, j}^t
 = \left\{\!\!
   \begin{array}{l@{~}l}
      1, & \mbox{if channel } j \mbox{ is allocated to VUE-pair } k     \\
         & \mbox{during decision epoch } t;                            \\
      0, & \mbox{otherwise}.
   \end{array}
   \right.
\end{align}
Thus we have
\begin{align}
  \sum_{j \in \mathcal{J}}   u_{k, j}^t & \leq 1, \forall k \in \mathcal{K},                                   \label{chanCons01}\\
  \sum_{k \in \mathcal{K}_i} u_{k, j}^t & \leq 1, \forall j \in \mathcal{J}, \forall i \in \mathcal{I}.        \label{chanCons02}
\end{align}
At the vTx of each VUE-pair $k$, a data queue is maintained to buffer the arriving packets.
Let $a_k^t$ be the random new packet arrivals at epoch $t$ with average arrival rate $\textsf{E}[a_k^t] = \lambda$.
The queue evolution for VUE-pair $k$ can be expressed as
\begin{align}\label{queuEvol}
    q_k^{t + 1} = \max\!\left\{q_k^t - r_k^t \cdot \mathds{1}_{\left\{\sum_{j \in \mathcal{J}} u_{k, j}^t = 1\right\}}, 0\right\} + a_k^t,
\end{align}
where $q_k^t$ and $r_k^t$ are, respectively, the queue length and the number of packets to depart during decision epoch $t$, while $\mathds{1}_{\{\Xi\}}$ is an indicator function that equals $1$ if the condition $\Xi$ is satisfied and $0$ otherwise.
In this paper, we assume a large enough buffer size to neglect the probability of packet drops.
The required transmit power for delivering $r_k^t \cdot \mathds{1}_{\left\{\sum_{j \in \mathcal{J}} u_{k, j}^t = 1\right\}}$ packets can be computed as
\begin{align}\label{poweCons}
  p_k^t = \dfrac{\vartheta + w \cdot \sigma^2}{g_{k, j}^t} \cdot \left(2^{\frac{\mu \cdot r_k^t}{w \cdot \delta}} - 1\right) \cdot
          \mathds{1}_{\left\{u_{k, j}^t = 1\right\}},
\end{align}
where $\vartheta$ is the received interference due to inter-group channel reuse, $w$ is the frequency bandwidth of the channels, $\sigma^2$ is the power spectral density of additive background noise, and $\mu$ is the constant size of a data packet.

\section{Problem Description}
\label{probForm}

This section formulates the problem of RRM in the considered V2V network as a discrete-time MDP with a discounted criterion and discusses the general solution.

\subsection{MDP Formulation}

During each decision epoch $t$, the local state of a VUE-pair $k \in \mathcal{K}$ can be described by $\mathbf{s}_k^t = (\mathbf{g}_k^t, (\mathbf{x}_k^t, \mathbf{y}_k^t), q_k^t) \in \mathcal{S} = \mathcal{G}^J \times \mathcal{C} \times \mathcal{Q}$, which includes the information of channel quality $\mathbf{g}_k^t = (g_{k, j}^t: j \in \mathcal{J})$, geographical location $(\mathbf{x}_k^t, \mathbf{y}_k^t)$ and queue state $q_k^t$.
We use $\mathbf{s}^t = (\mathbf{s}_k^t, \mathbf{s}_{-k}^t) \in \mathcal{S}^K$ to represent the global network state, where $-k$ denotes all the other VUE-pairs in $\mathcal{K}$ without the presence of VUE-pair $k$.
The RSU aims to design a stationary control policy $\bm\pi = (\pi_{(u)}, \pi_{(r)})$, where $\pi_{(u)}$ and $\pi_{(r)}$ are, respectively, the channel allocation policy and the packet scheduling policy.
Specifically, the RSU observes $\mathbf{s}^t$ at the beginning of epoch $t$ and accordingly, makes channel allocation and packet scheduling decisions for the VUE-pairs, that is, $\bm\pi(\mathbf{s}^t) = (\pi_{(u)}(\mathbf{s}^t), \pi_{(r)}(\mathbf{s}^t)) = (\mathbf{u}^t, \mathbf{r}^t)$, where $\mathbf{u}^t = (\mathbf{u}_k^t: k \in \mathcal{K})$ and $\mathbf{r}^t = (r_k^t: k \in \mathcal{K})$.
From the assumptions on the mobility of a VUE-pair, the packet arrivals and the queue evolution, the randomness lying in $\{\mathbf{s}^t: t \in \mathds{N}_+\}$ is Markovian with the following controlled state transition probability
\begin{align}\label{statTranProb}
 & \mathbb{P}\!\left(\mathbf{s}^{t + 1} | \mathbf{s}^t, \bm\pi\left(\mathbf{s}^t\right)\right)
 = \prod_{k \in \mathcal{K}} \mathbb{P}\!\left(g_k^{t + 1} | \left(\mathbf{x}_k^{t + 1}, \mathbf{y}_k^{t + 1}\right)\right) \cdot           \nonumber\\
 & \mathbb{P}\!\left(\left(\mathbf{x}_k^{t + 1}, \mathbf{y}_k^{t + 1}\right) | \left(\mathbf{x}_k^t, \mathbf{y}_k^t\right)\right) \cdot
   \mathbb{P}\!\left(q_k^{t + 1} | q_k^t, \mathbf{u}_k^t, r_k^t\right),
\end{align}
where $\mathbb{P}(\cdot)$ denotes the probability of an event.

We need a cost function to tradeoff the queuing delay and the consumed transmit power for each VUE-pair $k \in \mathcal{K}$ during each decision epoch $t$, which can be chosen as
\begin{align}\label{costFunc}
  f_k\!\left(\mathbf{s}^t, \mathbf{u}_k^t, r_k^t\right) = \phi \cdot d\!\left(q_k^t\right) + \eta \cdot p_k^t,
\end{align}
where $d(q_k^t) = q_k^t / \lambda$, while $\phi$ and $\eta$ are two positive weights.
Given a control policy $\bm\pi$ and an initial global network state $\mathbf{s}^1 = \mathbf{s} \in \mathcal{S}^K$, we express the expected long-term cost function $V_k(\mathbf{s}, \bm\pi)$ for VUE-pair $k$ as
\begin{align}\label{statValu}
  V_k\!\left(\mathbf{s}, \bm\pi\right) = (1 - \gamma) \cdot
  \textsf{E}_{\bm\pi}\!\!\left[\sum_{t = 1}^\infty (\gamma)^{t  - 1} f_k\!\left(\mathbf{s}^t, \mathbf{u}_k^t, r_k^t\right) | \mathbf{s}\right],
\end{align}
where $\gamma \in [0, 1)$ is the discount factor.
As a result, the delay-power tradeoff problem, which the RSU aims to solve, can be formally formulated as a MDP, namely, $\forall \mathbf{s} \in \mathcal{S}^K$,
\begin{align}\label{MDP}
 \min_{\bm\pi} V(\mathbf{s}, \bm\pi) & = \displaystyle\sum_{k \in \mathcal{K}} V_k\!\left(\mathbf{s}, \bm\pi\right)                         \\
                                     & = (1 - \gamma) \cdot \textsf{E}_{\bm\pi}\!\!\left[\displaystyle\sum_{t = 1}^\infty (\gamma)^{t  - 1}
                                         f\!\left(\mathbf{s}^t, \bm\pi(\mathbf{s}^t)\right) | \mathbf{s}\right]                             \nonumber\\
 \mathrm{s.t.}~~~~~~~~~~             &~  \mbox{constraints } (\ref{chanCons01}) \mbox{ and } (\ref{chanCons02}),                            \nonumber
\end{align}
where $f(\mathbf{s}^t, \bm\pi(\mathbf{s}^t)) = \sum_{k \in \mathcal{K}} f_k(\mathbf{s}^t, \mathbf{u}_k^t, r_k^t)$ is the immediate cost accumulated across all the VUE-pairs in the network at a decision epoch $t$.
$V(\mathbf{s}, \bm\pi)$ is also named as the state value function in state $\mathbf{s}$ under a policy $\bm\pi$.

\subsection{Optimal Solution}
\label{optiSolu}

The problem formulated as in (\ref{MDP}) is a typical infinite-horizon discrete-time MDP with a discounted criterion.
Denote by $\bm\pi^* = (\pi_{(u)}^*, \pi_{(r)}^*)$ the optimal control policy, which can be obtained from solving the Bellman's equation: $\forall \mathbf{s} \in \mathcal{S}^K$,
\begin{align}\label{BellEqua}
 & V(\mathbf{s}) =                                                                                                                               \\
 & \min_{\bm\pi(\mathbf{s})}\!\left\{(1 - \gamma) \cdot f(\mathbf{s}, \bm\pi(\mathbf{s})) +
   \gamma \cdot \! \sum_{\mathbf{s}' \in \mathcal{S}^K} \! \mathbb{P}(\mathbf{s}' | \mathbf{s},\bm\pi(\mathbf{s})) \cdot V(\mathbf{s}')\right\}, \nonumber
\end{align}
where $V(\mathbf{s}) = V(\mathbf{s}, \bm\pi^*)$ is the optimal state value function and $\mathbf{s}' \in \mathcal{S}^K$ is the resulting global network state at a subsequent epoch.
The conventional solutions to (\ref{BellEqua}) based on the value or policy iteration \cite{Rich98} require the complete knowledge of network dynamics (\ref{statTranProb}), which is challenging in practice.
Let us define the right-hand side of (\ref{BellEqua}) by
\begin{align}\label{stat_acti_q1}
     Q(\mathbf{s}, \mathbf{u}, \mathbf{r})
 & = (1 - \gamma) \cdot f(\mathbf{s}, \mathbf{u}, \mathbf{r})                                                                               \nonumber\\
 & + \gamma \cdot \sum_{\mathbf{s}' \in \mathcal{S}^K} \mathbb{P}(\mathbf{s}' | \mathbf{s}, \mathbf{u}, \mathbf{r}) \cdot V(\mathbf{s}'),
\end{align}
the $Q$-function, where $\mathbf{u} = (\mathbf{u}_k: k \in \mathcal{K})$ and $\mathbf{r} = (r_k: k \in \mathcal{K})$ are the decision makings under $\mathbf{s}$ with $\mathbf{u}_k = (u_{k, j}: j \in \mathcal{J})$.
$V(\mathbf{s})$ can then be directly obtained from
\begin{align}\label{stat_acti_q2}
    V(\mathbf{s}) = \min_{\mathbf{u}, \mathbf{r}} Q(\mathbf{s}, \mathbf{u}, \mathbf{r}).
\end{align}
By substituting (\ref{stat_acti_q2}) back into (\ref{stat_acti_q1}), we have
\begin{align}
     Q(\mathbf{s}, \mathbf{u}, \mathbf{r})
 & = (1 - \gamma) \cdot f(\mathbf{s}, \mathbf{u}, \mathbf{r})                                                               \nonumber\\
 & + \gamma \cdot \sum_{\mathbf{s}' \in \mathcal{S}^K} \mathbb{P}(\mathbf{s}' | \mathbf{s}, \mathbf{u}, \mathbf{r}) \cdot
     \min_{\mathbf{u}', \mathbf{r}'} Q(\mathbf{s}', \mathbf{u}', \mathbf{r}'),
\end{align}
where $\mathbf{u}' = (\mathbf{u}_k': k \in \mathcal{K})$ and $\mathbf{r}' = (r_k': k \in \mathcal{K})$ denote the decision makings under $\mathbf{s}'$ with $\mathbf{u}_k' = (u_{k, j}': j \in \mathcal{J})$.

Using a state-action-reward-state-action (SARSA) algorithm \cite{Rumm94, Rich98}, the RSU tries to learn $Q(\mathbf{s}, \mathbf{u}, \mathbf{r})$ in a recursive way with observations of the global network state $\mathbf{s} = \mathbf{s}^t$, the decision making $(\mathbf{u}, \mathbf{r}) = (\mathbf{u}^t, \mathbf{r}^t)$, the realized cost $f(\mathbf{s}, \mathbf{u}, \mathbf{r})$ at a current decision epoch $t$ and the resulting global network state $\mathbf{s}' = \mathbf{s}^{t + 1}$, the decision making $(\mathbf{u}', \mathbf{r}') = (\mathbf{u}^{t + 1}, \mathbf{r}^{t + 1})$ at the next epoch $t + 1$.
The updating rule is given by
\begin{align}\label{QLearRule}
 & Q^{t + 1}(\mathbf{s}, \mathbf{u}, \mathbf{r}) = Q^t(\mathbf{s}, \mathbf{u}, \mathbf{r}) + \\
 & \alpha^t \cdot \left((1 - \gamma) \cdot f(\mathbf{s}, \mathbf{u}, \mathbf{r}) + \gamma \cdot Q^t(\mathbf{s}', \mathbf{u}', \mathbf{r}') -
   Q^t(\mathbf{s}, \mathbf{u}, \mathbf{r})\right),\nonumber
\end{align}
where $\alpha^t \in [0, 1)$ is the learning rate.
It has been proven that if 1) the network state transition probability under the optimal stationary control policy is stationary, 2) $\sum_{t = 1}^\infty \alpha^t$ is infinite and $\sum_{t = 1}^\infty (\alpha^t)^2$ is finite, and 3) all state-action pairs are visited infinitely often (which can be satisfied by a $\epsilon$-greedy strategy \cite{Rich98}), the SARSA learning process converges and finds $\bm\pi^*$ \cite{Sing00}.
However, two challenges remain as follows:
\begin{enumerate}
  \item from the channel model applied in this work, the global network state space $\mathcal{S}^K$ is \emph{semi-continuous}; and
  \item the number $((1 + J) \cdot (1 + A))^K$ of decision makings at the RSU grows \emph{exponentially} as $K$ increases, where $A$ is the maximum number of packet departures at a vTx, i.e., $a_k^t \leq A$, $\forall k \in \mathcal{K}$ and $\forall t \in \mathds{N}_+$.
\end{enumerate}

\section{A Deep Reinforcement Learning Approach}
\label{probSolv}

We shall address in this section the technical challenges in solving an optimal control policy and derive a deep reinforcement learning algorithm.

\subsection{Linear $Q$-function Decomposition}

The centralized decisions made by the RSU are performed by the VUE-pairs in a decentralized way.
We hence propose to linearly decompose the $Q$-function, that is,
\begin{align}\label{QFuncDeco}
   Q(\mathbf{s}, \mathbf{u}, \mathbf{r}) = \sum_{k \in \mathcal{K}} Q_k(\mathbf{s}, \mathbf{u}_k, r_k),
\end{align}
where $Q_k(\mathbf{s}, \mathbf{u}_k, r_k)$ is the per-VUE-pair $Q$-function for each VUE-pair $k \in \mathcal{K}$ that satisfies
\begin{align}\label{perVUEPQFunc}
 & Q_k(\mathbf{s}, \mathbf{u}_k, r_k) =
   (1 - \gamma) \cdot f_k(\mathbf{s}, \mathbf{u}_k, r_k) +                                                               \\
 & \gamma \cdot \sum_{\mathbf{s}' \in \mathcal{S}^K}
   \mathbb{P}\!\left(\mathbf{s}' | \mathbf{s}, (\mathbf{u}_k, \mathbf{u}_{-k}), (r_k, \mathbf{r}_{-k})\right) \cdot
   Q_k(\mathbf{s}', \mathbf{u}_k', r_k'),                                                                                \nonumber
\end{align}
where the optimal decision making from a VUE-pair $k$ across the time horizon should reflect the optimal control policy implemented by the RSU.
In other words, $(\mathbf{u}_k', r_k')$ in (\ref{perVUEPQFunc}) under the network state $\mathbf{s}'$ follows $\bm\pi^*(\mathbf{s}')$, i.e.,
\begin{align}
  \bm\pi^*(\mathbf{s}') = \underset{\mathbf{u}', \mathbf{r}'}{\arg\min} \sum_{k \in \mathcal{K}} Q_k(\mathbf{s}', \mathbf{u}_k', r_k'),
\end{align}
which minimizes the sum of per-VUE-pair $Q$-function values from all VUE-pairs in the network.
Two key advantages of the decomposition approach in (\ref{QFuncDeco}) are highlighted.
\begin{enumerate}
  \item Simplified decision makings: The linear decomposition motivates the RSU to let the VUE-pairs submit the local per-VUE-pair $Q$-functions of the channel allocation and packet scheduling decisions with the global network state observations, based on which the RSU allocates channels and the VUE-pairs then schedule packet transmissions.
      This reduces $((1 + J) \cdot (1 + A))^K$ centralized decision makings at the RSU to $K \cdot ((1 + J) \cdot (1 + A))$ decentralized decisions for all VUE-pairs.
  \item Near optimality: The approach in (\ref{QFuncDeco}) ensures a guarantee of approximation error of the $Q$-function \cite{Chen1804}.
\end{enumerate}

\subsection{Learning the Optimal Control Policy}

In spite of the advantages brought by the linear decomposition approach as in (\ref{QFuncDeco}), a new challenge, however, arises.
That is, each VUE-pair $k \in \mathcal{K}$ can only obtain a partial observation $(\mathbf{s}_k^t, \mathbf{o}_k^t)$ of the global network state $\mathbf{s}^t$ at each decision epoch $t$.
In this work, we assume that when VUE-pair $k$ was in a group $i_k^{t - 1} \in \mathcal{I}$ (i.e., $k \in \mathcal{K}_{i_k^{t - 1}}$) during the previous decision epoch $t - 1$, $\mathbf{o}_k^t = (i_k^{t - 1}, b_{i_k^{t - 1}}^{t - 1}, \bm\upsilon_{i_k^{t - 1}}^{t - 1}) \in \mathcal{O}$ includes the group index $i_k^{t - 1}$ and the number $b_{i_k^{t - 1}}^{t - 1}$ of VUE-pairs as well as the channel utilization state $\bm\upsilon_{i_k^{t - 1}}^{t - 1} = (\upsilon_{i_k^{t - 1}, j}^{t - 1}: j \in \mathcal{J})$ in group $i_k^{t - 1}$, where $\upsilon_{i_k^{t - 1}, j}^{t - 1}$ equals $1$ if channel $j \in \mathcal{J}$ is utilized in group $i_k^{t - 1}$ at epoch $t - 1$ and otherwise, $0$.
Note that $\mathbf{o}_k^t$ is restricted to local group information since the decision makings across different groups are independent.

\begin{figure}[t]
  \centering
  \includegraphics[width=18pc]{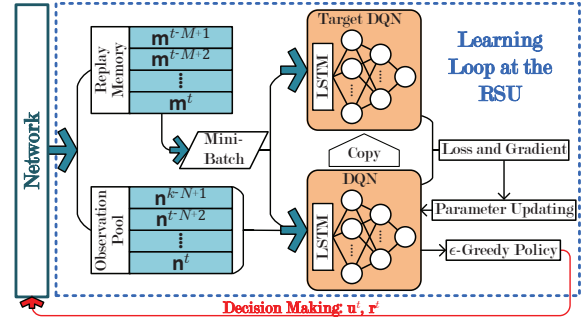}
  \caption{Long short-term memory (LSTM) based deep reinforcement learning for long-term delay-power tradeoff in a vehicle-to-vehicle network (RSU: road side unit; DQN: deep $Q$-network.).}
  \label{deepLear}
\vspace{-0.5cm}
\end{figure}

With the local observation $(\mathbf{s}_k, \mathbf{o}_k) \in \mathcal{S} \times \mathcal{O}$ at a current decision epoch, we abstract the per-VUE-pair $Q$-function (\ref{perVUEPQFunc}) of each VUE-pair $k \in \mathcal{K}$ as \cite{Chen1804}
\begin{align}\label{qFuncObse}
  Q_k(\mathbf{s}, \mathbf{u}_k, r_k) \approx Q_k(\mathbf{s}_k, \mathbf{o}_k, \mathbf{u}_k, r_k).
\end{align}
The semi-continuity in $\mathcal{S}$ and the high dimensionality in $\mathcal{O}$ make it infeasible for the conventional SARSA algorithm (\ref{QLearRule}) to learn the per-VUE-pair $Q$-function $Q_k(\mathbf{s}_k, \mathbf{o}_k, \mathbf{u}_k, r_k)$, $\forall k \in \mathcal{K}$.
Moreover, from the assumptions made in this paper and the definition of a cost function (\ref{costFunc}), there exists homogeneity in the VUE-pair behaviours.
Inspired by the success of modelling the $Q$-function with a deep neural network (DNN) \cite{Mnih15}, we adopt a common double deep $Q$-network (DQN) to approximate $Q_k(\mathbf{s}_k, \mathbf{o}_k, \mathbf{u}_k, r_k)$ \cite{Hass16, Dai19}.
On the other hand, the accuracy of (\ref{qFuncObse}) from the observations can be, in general, arbitrarily bad.
As in \cite{Hauk15}, we propose to add a LSTM layer \cite{Hoch97} to the DQN and obtain a hybrid DNN to learn a better control policy in a partially observable V2V network.
Specifically, let $Q_k(\mathbf{s}, \mathbf{u}_k, r_k)$ $\approx Q_k(\mathcal{N}_k, \mathbf{u}_k, r_k; \bm\theta)$, $\forall k \in \mathcal{K}$, where $\mathcal{N}_k$ denotes a set of most recent $N$ local observations up to a current decision epoch (which will be specified later in this subsection) and is taken as an input to the LSTM layer for a more accurate prediction of $\mathbf{s}$, while $\bm\theta$ denotes a vector of parameters associated with the hybrid DNN.
Our proposed novel LSTM based deep reinforcement learning (LSTM-DRL) algorithm for long-term delay-power tradeoff in the considered V2V network is illustrated in Fig. \ref{deepLear}, during which instead of finding the per-VUE-pair $Q$-function, the parameters of the hybrid DNN can be trained centrally at the RSU.

For online training of the LSTM-DRL algorithm, at each decision epoch $t$, the RSU updates the replay memory $\mathcal{M}$ with the most recent $M$ experiences $\{\mathbf{m}^{t - M + 1}, \cdots, \mathbf{m}^t\}$ with each experience $\mathbf{m}^{t - m + 1}$ ($\forall m \in \{1, \cdots, M\}$) being given by
\begin{align}\label{expe}
 & \mathbf{m}^{t - m + 1} =                                                                                         \nonumber\\
 & \left(\left(\left(\mathbf{s}_k^{t - m}, \mathbf{o}_k^{t - m}\right),
         \left(\mathbf{u}_k^{t - m}, r_k^{t - m}\right),
         f_k\!\left(\mathbf{s}^{t - m}, \mathbf{u}_k^{t - m}, r_k^{t - m}\right),\right.\right.                     \nonumber\\
 & \left.\left.~~\left(\mathbf{s}_k^{t - m + 1}, \mathbf{o}_k^{t - m + 1}\right),
         \left(\mathbf{u}_k^{t - m + 1}, r_k^{t - m + 1}\right)\right): k \in \mathcal{K}\right).
\end{align}
Meanwhile, an observation pool $\mathcal{N}^t = \cup_{k \in \mathcal{K}} \mathcal{N}_k^t = \{\mathbf{n}^{t - N + 1},$ $\cdots, \mathbf{n}^{t}\}$, the information of which is collected from all VUE-pairs, is kept to predict the global network state $\mathbf{s}^t$ at epoch $t$ for control policy evaluation, where $\mathbf{n}^{t} = \{\mathbf{n}_k^{t} = (\mathbf{s}_k^t, \mathbf{o}_k^t): k \in \mathcal{K}\}$.
To train the hybrid DNN parameters, the RSU first randomly samples a mini-batch $\widetilde{\mathcal{M}}^t = \{\widetilde{\mathcal{M}}^{t_1}, \cdots, \widetilde{\mathcal{M}}^{t_{\widetilde{M}}}\}$ of size $\widetilde{M}$ from $\mathcal{M}^t$, where $\forall m \in \{1, \cdots, \widetilde{M}\}$,
\begin{align}
   \widetilde{\mathcal{M}}^{t_m} =
 & \left\{\left(\mathcal{N}_k^{t_m}, \left(\mathbf{u}_k^{t_m}, r_k^{t_m}\right),
          f_k\!\left(\mathbf{s}^{t_m}, \mathbf{u}_k^{t_m}, r_k^{t_m}\right), \right.\right.                                         \nonumber\\
 & \left.\left.~~\mathcal{N}_k^{t_m + 1}, \left(\mathbf{u}_k^{t_m + 1}, r_k^{t_m + 1}\right)\right): k \in \mathcal{K}\right\},
\end{align}
with $\mathcal{N}_k^{t_m} = \{\mathbf{n}_k^{t_m - N + 1}, \cdots, \mathbf{n}_k^{t_m}\}$.
Then the set $\bm\theta^t$ of parameters at epoch $t$ is updated by minimizing the accumulative loss function, which is defined as in (\ref{lossFunc}),
\begin{figure*}[!t]
\vspace{-0.2cm}
\begin{align}\label{lossFunc}
 & L\!\left(\bm\theta^t\right) =                                                                                                    \nonumber\\
 & \textsf{E}_{\left\{\left(\left(\mathcal{N}_k, (\mathbf{u}_k, r_k), f_k(\mathbf{s}, \mathbf{u}_k, r_k),
                      \mathcal{N}_k', \left(\mathbf{u}_k', r_k'\right)\right): k \in \mathcal{K}\right) \in \widetilde{\mathcal{M}}^t\right\}}\!
   \left[\left(\sum_{k \in \mathcal{K}}
   \left(\!\!
   \begin{array}{c}
      \displaystyle(1 - \gamma) \cdot \displaystyle f_k(\mathbf{s}, \mathbf{u}_k, r_k) + \gamma \cdot
            Q_k\!\left(\mathcal{N}_k', \mathbf{u}_k', r_k'; \bm\theta_{-}^t\right) -                                  \\
      Q_k\!\left(\mathcal{N}_k, \mathbf{u}_k, r_k; \bm\theta^t\right)
   \end{array}\!\!
   \right)
   \right)^2\right]
\end{align}
\vspace{-0.2cm}
\hrule
\vspace{-0.5cm}
\end{figure*}
where $\bm\theta_{-}^t$ is the set of parameters of the target hybrid DNN at a certain previous decision epoch before epoch $t$.
The gradient is calculated as (\ref{grad}).
\begin{figure*}[!t]
\vspace{-0.2cm}
\begin{align}\label{grad}
 & \nabla_{\bm\theta^t} L\!\left(\bm\theta^t\right) =                                                                               \nonumber\\
 & \textsf{E}_{\left\{\left(\left(\mathcal{N}_k, (\mathbf{u}_k, r_k), f_k(\mathbf{s}, \mathbf{u}_k, r_k),
                      \mathcal{N}_k', \left(\mathbf{u}_k', r_k'\right)\right): k \in \mathcal{K}\right) \in \widetilde{\mathcal{M}}^t\right\}}\!
   \left[\!\!
   \begin{array}{c}
     \displaystyle\sum_{k \in \mathcal{K}}
       \left(\!\!
       \begin{array}{c}
          \displaystyle(1 - \gamma) \cdot \displaystyle f_k(\mathbf{s}, \mathbf{u}_k, r_k) + \gamma \cdot
              Q_k\!\left(\mathcal{N}_k', \mathbf{u}_k', r_k'; \bm\theta_{-}^t\right) -                                \\
          Q_k\!\left(\mathcal{N}_k, \mathbf{u}_k, r_k; \bm\theta^t\right)
       \end{array}\!\!
       \right) \cdot                                                                                                                \\
     \nabla_{\bm\theta^t}\!\! \left(\displaystyle\sum_{k \in \mathcal{K}}
       Q_k\!\left(\mathcal{N}_k, \mathbf{u}_k, r_k; \bm\theta^t\right)\right)
   \end{array}\!\!
   \right]
\end{align}
\vspace{-0.2cm}
\hrule
\vspace{-0.5cm}
\end{figure*}
We summarize in Algorithm \ref{algo} the online training of the proposed LSTM-DRL algorithm.

\begin{algorithm}[t]
    \caption{Online Training of LSTM-DRL for Long-Term Delay-Power Tradeoff in V2V Networks}
    \label{algo}
    \begin{algorithmic}[1]
        \STATE \textbf{initialize} the replay memory $\mathcal{M}^t$ with size $M$, the observation pool $\mathcal{N}^t$ with size $N$, the mini-batch $\widetilde{\mathcal{M}}^t$ with size $\widetilde{M}$ and the decision making $(\mathbf{u}^t, \mathbf{r}^t)$, for $t = 1$.

        \REPEAT
            \STATE After performing $(\mathbf{u}^t, \mathbf{r}^t)$ at epoch $t$, each VUE-pair $k \in \mathcal{K}$ realizes an immediate cost $f_k(\mathbf{s}^t, \mathbf{u}_k^t, r_k^t)$ .

            \STATE Each VUE-pair $k$ observes $(\mathbf{s}_k^{t + 1}, \mathbf{o}_k^{t + 1}) \in \mathcal{S} \times \mathcal{O}$ at the next decision epoch $t + 1$.

            \STATE The RSU updates the observation pool $\mathcal{N}^t$ with $\mathbf{n}^{t + 1} = \{(\mathbf{s}_k^{t + 1}, \mathbf{o}_k^{t + 1}): k \in \mathcal{K}\}$ collected from all VUE-pairs.

            \STATE With probability $\epsilon$, the RSU selects a decision making $(\mathbf{u}^{t + 1}, \mathbf{r}^{t + 1})$ randomly; or with probability $1 - \epsilon$, the RSU takes $\mathcal{N}^{t + 1}$ as the input to the hybrid DNN with parameters $\bm\theta^t$, and then determines $(\mathbf{u}^{t + 1}, \mathbf{r}^{t + 1}) = \arg\min_{\mathbf{u}, \mathbf{r}} \sum_{k \in \mathcal{K}} Q_k(\mathcal{N}_k^{t + 1}, \mathbf{u}_k, r_k; \bm\theta^t)$.

            \STATE The RSU updates the replay memory $\mathcal{M}^{t + 1}$ with the most recent experience $\mathbf{m}^{t + 1}$ in the form of (\ref{expe}).

            \STATE With a randomly sampled mini-batch $\widetilde{\mathcal{M}}^t$ from $\mathcal{M}^t$, the RSU updates the hybrid DNN parameters $\bm\theta^t$ with the gradient given by (\ref{grad}).

            \STATE The RSU regularly resets the target DQN with parameters $\bm\theta_{-}^{t + 1}$ with $\bm\theta^t$, and otherwise, $\bm\theta_{-}^t$.

            \STATE The decision epoch index is updated by $t \leftarrow t + 1$.
        \UNTIL{A predefined stopping condition is satisfied.}
    \end{algorithmic}
\end{algorithm}

\section{Simulation Results}
\label{simuResu}

This section evaluates the performance from our proposed studies through numerical simulations based on TensorFlow \cite{Abad16}.
We simulate a $250\times250$ m$^2$ Manhattan mobility model with nine intersections \cite{Liu18, Chen18S}.
In the model, a road consists of two lanes, each of which is in one direction and is of width $4$ m.
The average vehicle speed is set to be $60$ km/h, and the vehicle grouping is performed by means of spectral clustering \cite{Chen18S}.
We list other parameter values used in simulations in Table \ref{tabl1}.
For performance comparison purpose, the following three baseline algorithms are simulated as well.
\begin{enumerate}
  \item \emph{Channel-Aware:} At each decision epoch, the RSU allocates the channels to VUE-pairs in each group based on the channel quality states.
  \item \emph{Queue-Aware:} Different from the Channel-Aware algorithm, the RSU allocates at each decision epoch the channels to VUE-pairs in each group according to the queue lengths.
  \item \emph{Random:} Across the decision epochs, the RSU randomly allocates the channels to a set of randomly picked VUE-pairs in each group.
\end{enumerate}
Implementing these baselines, the RSU schedules packets to minimize the immediate cost for each VUE-pair.

\begin{table}[t]
  \caption{Parameter values in simulations.}\label{tabl1}
        \begin{center}
        \begin{tabular}{c|c}
              \hline
              Parameter                                     & Value                                     \\\hline
              Replay memory capacity        $M$             & $5000$                                    \\\hline
              Mini-batch size               $\tilde{M}$     & $200$                                     \\\hline
              Observation pool size         $N$             & $20$                                      \\\hline
              Path loss exponent $\rho$, $\xi$              & $-68.5$ dB, $-54.5$ dB                    \\\hline
              Path loss coefficient $e$                     & $1.61$                                    \\\hline
              Distance $\varphi_0$                          & $15$ m                                    \\\hline
              Number of VUE-pair group $I$                  & $10$                                      \\\hline
              Clustering interval $T$                       & $10$ epochs                               \\\hline
              Frequency bandwidth $w$                       & $500$ kHz                                 \\\hline
              Aggregate interference $\vartheta$            & $2 \cdot 10^{-9}$ W                       \\\hline
              Noise power spectral density $\sigma^2$       & $7.95 \cdot 10^{-21}$ W/Hz                \\\hline
              Scheduling epoch duration $\delta$            & $18$ ms                                   \\\hline
              Weights $\phi$, $\eta$                        & $30$, $1$                                     \\\hline
              Data packet size $\mu$                        & $9$ kb                                    \\\hline
              Discount factor  $\gamma$                     & $0.9$                                     \\\hline
              Exploration probability $\epsilon$            & $0.06$                                    \\
              \hline
        \end{tabular}
    \end{center}
\end{table}

\subsection{Convergence Property of the Proposed Algorithm}

\begin{figure}[t]
  \centering
  \includegraphics[width=13.6pc]{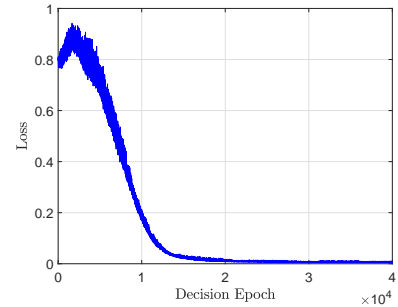}
  \caption{Illustration of the convergence property of our proposed online LSTM-DRL algorithm.}
  \label{sim01}
\vspace{-0.5cm}
\end{figure}

This simulation examines the convergence property of online training of our LSTM-DRL algorithm.
We select $K = 36$ VUE-pairs with an average packet arrival rate $\lambda = 1$, and the distance between the VTx and the vRx of each VUE-pair is fixed to be $\varphi = 20$.
Fig. \ref{sim01} plots the loss function defined by (\ref{lossFunc}) over the learning time horizon, which validates that the convergence needs around $3 \cdot 10^4$ decision epochs.
Since the training is performed centrally at the RSU, each VUE-pair only needs to periodically update the set $\bm\theta$ of parameters of the LSTM-DRL algorithm with a new one from the RSU.

\begin{figure*}
  \centering
  \subfigure[Average cost per VUE-pair versus number of VUE-pairs $K$: $\varphi = 20$ and $\lambda = 2$.]{\label{sim2}\includegraphics[width=13.6pc]{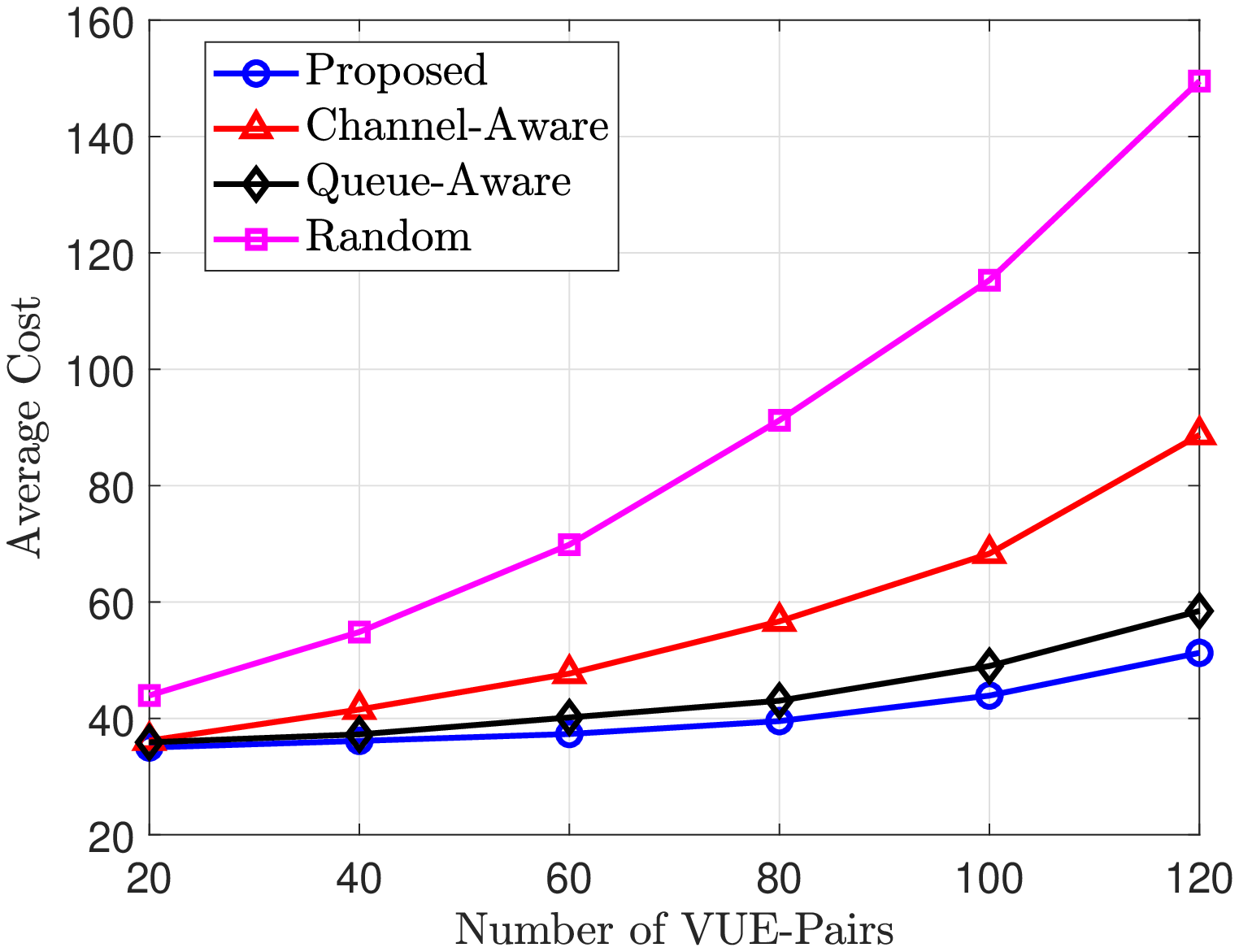}}
  \subfigure[Average cost per VUE-pair versus average packet arrival rate $\lambda$: $K = 52$ and $\varphi = 35$.]{\label{sim3}\includegraphics[width=13.6pc]{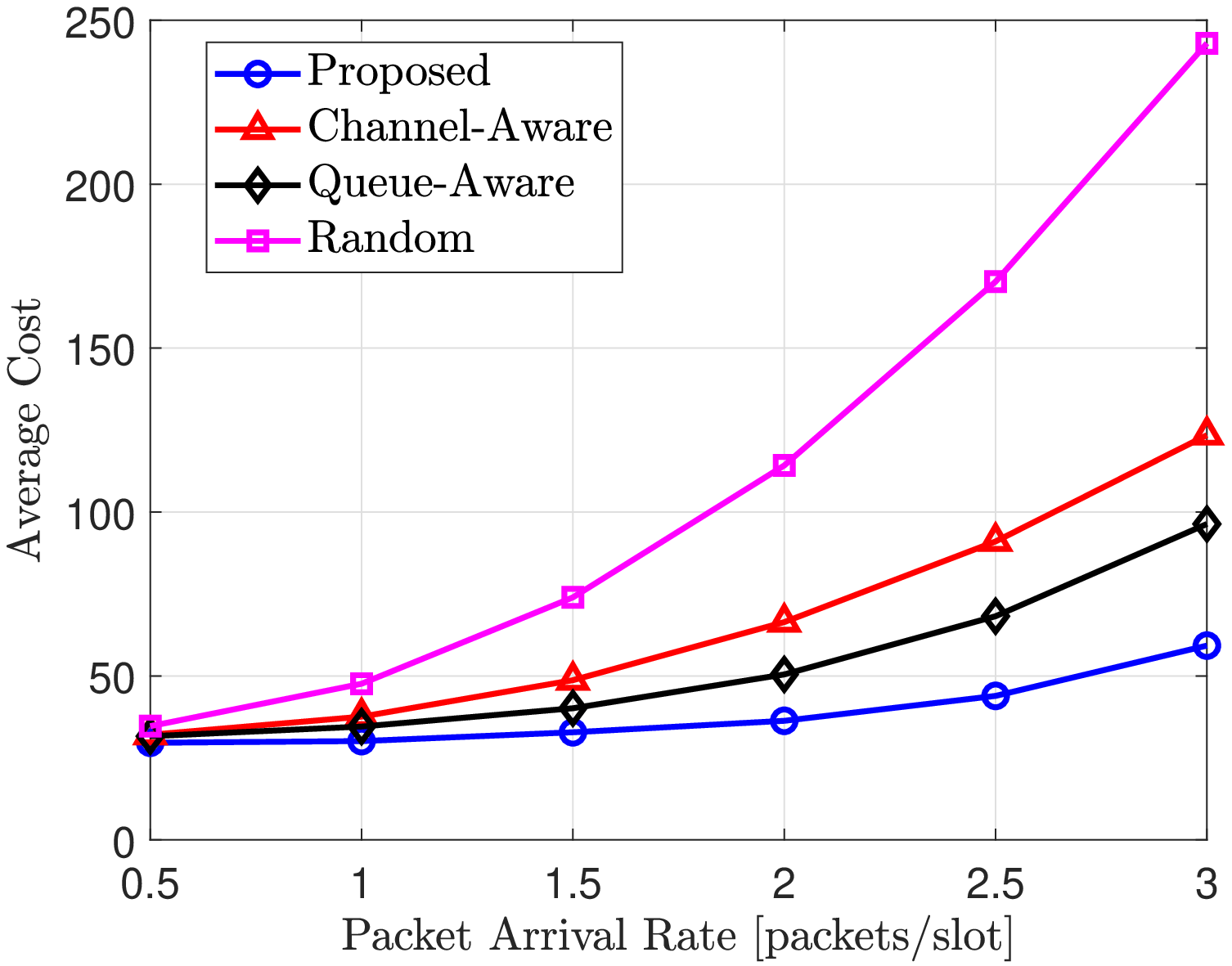}}
  \subfigure[Average cost per VUE-pair versus VUE-pair distance $\varphi$: $K = 36$ and $\lambda = 1$.]{\label{sim4}\includegraphics[width=13.6pc]{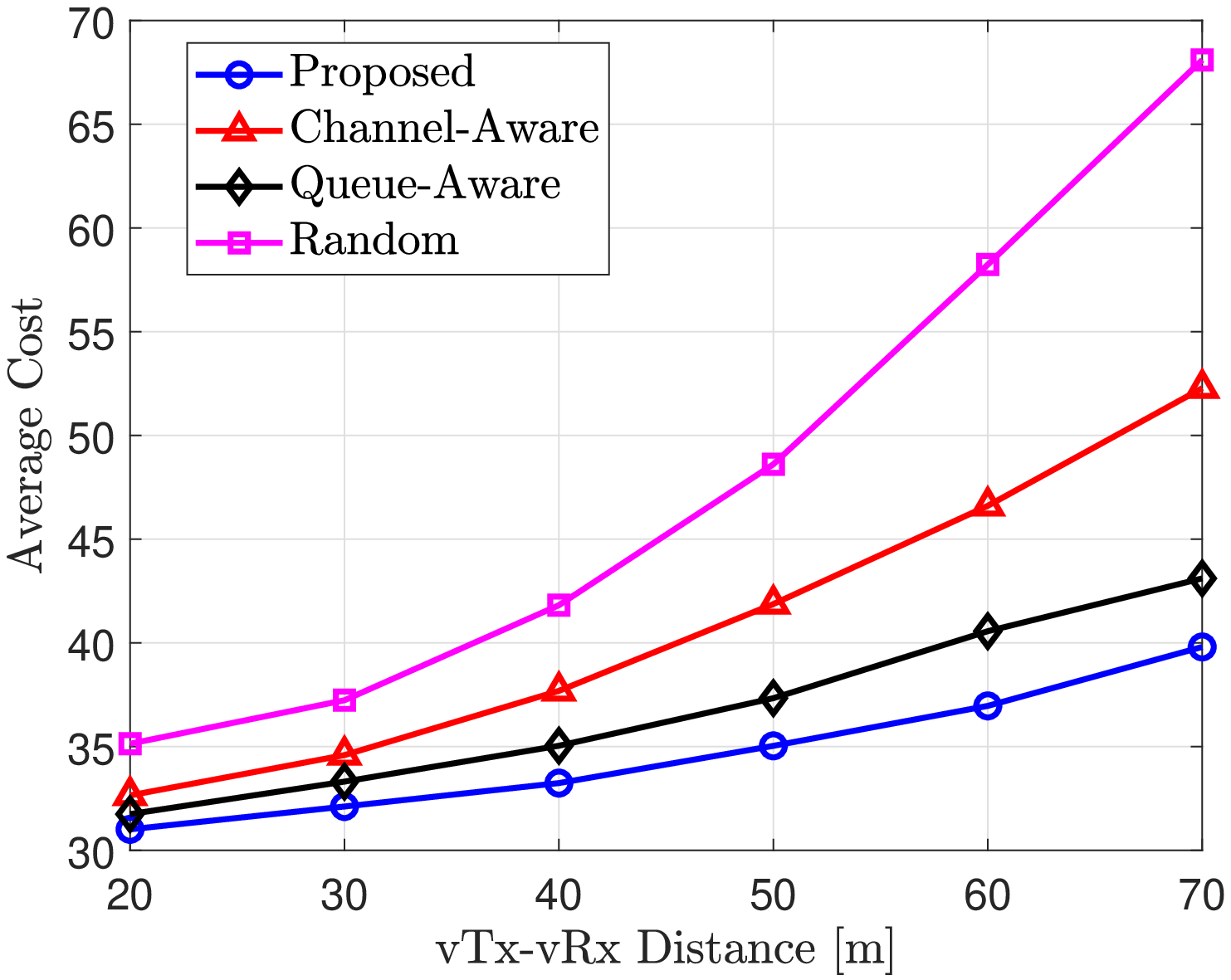}}
  \caption{Average cost performance per VUE-pair under various simulation settings.}\label{perf02}
\vspace{-0.5cm}
\end{figure*}

\subsection{Performance under Various Simulation Settings}
\label{simu02}

We further verify the average cost performance per VUE-pair across the time horizon under different simulation settings.
First, we configure a networking environment as: $\lambda = 2$ and $\varphi = 20$.
In Fig. \ref{sim2}, we depict the realized average cost performance versus $K$, which shows the average cost per VUE-pair from all four algorithms increases as the number of VUE-pairs increases.
It is obvious that a larger number of VUE-pairs leads to less chance of being allocated one channel.
Next, we assume there are $K = 52$ VUE-pairs in the network and $\varphi = 35$.
By increasing the value of $\lambda$, the average cost performance per VUE-pair is shown in Fig. \ref{sim3}
With more packets arriving into the queues, more power is consumed for the packet transmissions in order to maintain the queue stability.
Hence all four algorithms exhibit worse performance.
Finally, we illustrate in Fig. \ref{sim4} the average cost performance per VUE-pair when the value of $\varphi$ varies.
As the distance between the vTx and the vRx of a VUE-pair increases, the channel quality drops.
This indicates more transmit power for transmitting the same number of packets, which conforms what we see from the curves in Fig. \ref{sim4}.
Interestingly and importantly, in all above three simulations, our proposed algorithm achieves the best performance, demonstrating the feasibility of a better delay-power tradeoff, compared with the other three baselines.

\section{Conclusions}
\label{conc}

In this paper, we put our emphasis on investigating the RRM for an expected long-term delay-power tradeoff in a V2V communication network.
The RSU allocates channels and schedules packet transmissions for all VUE-pairs according to the observations of global network states over the discrete time horizon.
This kind of decision-making process straightforwardly falls into the realm of a MDP.
The technical challenges in solving an optimal control policy for the MDP motivates us to first decompose the MDP into a series of per-VUE-pair MDPs with much simplified decision makings.
To overcome the curse of high dimensionality in state space of a per-VUE-pair MDP, we resort to the DQN technique and propose an online LSTM-DRL algorithm.
The LSTM-DRL algorithm enables decentralized channel allocation and packet scheduling decisions with only partially local network state observations from the VUE-pairs but without a priori statistics knowledge of network dynamics.
From numerical simulations, significant gains in average cost performance from the proposed learning algorithm can be expected.

\end{document}